\documentclass{article}

\usepackage{makeidx}
\usepackage{physics}
\usepackage{xcolor}

\makeindex

\usepackage[sorting=none, backend=biber, style=numeric, maxbibnames=20]{biblatex}
\usepackage[english]{babel}
\usepackage{times}
\usepackage[T1]{fontenc} 
\usepackage{subcaption}
\usepackage{amsfonts, amsmath, multirow}
\usepackage{hyperref}
\newlength\LL 

\usepackage{graphicx}
\usepackage{array}
\usepackage{listings}
\usepackage{xfrac}
\usepackage{xspace}
\usepackage{colortbl}
\usepackage{alltt}
\usepackage{adjustbox} 
\usepackage{lcg}       

\usepackage{mathtools}       
\usepackage[ruled,noline,linesnumbered]{algorithm2e}
\usepackage{algpseudocode}   

\usepackage{tikz}
\usepackage{pgflibraryshapes}
\usetikzlibrary{decorations.text}
\usetikzlibrary{decorations.pathreplacing}
\usetikzlibrary{backgrounds}
\usetikzlibrary{calc}
\usetikzlibrary{arrows}
\usetikzlibrary{snakes}
\usetikzlibrary{positioning}
\usetikzlibrary{tikzmark}
\usepackage{pgfplots}
\usepackage{pgfplotstable}

\usepackage[absolute,overlay]{textpos}
\textblockorigin{0pt}{0pt} 

\makeatletter
\DeclareRobustCommand\onedot{\futurelet\@let@token\@onedot}
\def\@onedot{\ifx\@let@token.\else.\null\fi\xspace}

\makeatother

\makeatletter
\def\pd{\@ifnextchar[{\@pdwith}{\@pdwithout}}
\def\@pdwith[#1]#2#3{\href{http://www.mcs.anl.gov/petsc/petsc-3.7/docs/manualpages/#2/#1.html}{#3}}
\def\@pdwithout#1#2{\@pdwith[#2]{#1}{#2}}
\makeatother

\DeclareFixedFont{\ttb}{T1}{txtt}{bx}{n}{12} 
\DeclareFixedFont{\ttm}{T1}{txtt}{m}{n}{12}  

\usepackage{color}
\definecolor{deepblue}{rgb}{0,0,0.5}
\definecolor{deepred}{rgb}{0.6,0,0}
\definecolor{deepgreen}{rgb}{0,0.5,0}
\definecolor{riceGrey}{RGB}{94,96,98}

\newcommand\pythonstyle{\lstset{
language=Python,
basicstyle=\ttm\scriptsize,
otherkeywords={self},             
keywordstyle=\ttb\scriptsize\color{deepblue},
emph={MyClass,__init__},          
emphstyle=\ttb\scriptsize\color{deepred},    
commentstyle=\scriptsize\color{deepgreen},
frame=tb,                         
showstringspaces=false            %
}}

\lstnewenvironment{python}[1][]
{
\pythonstyle
\lstset{#1}
}
{}

\newcommand\pythoninline[1]{{\pythonstyle\lstinline!#1!}}

\newcommand\cstyle{\lstset{
language=C,
basicstyle=\ttm\scriptsize,
otherkeywords={MPI_Comm,TS,SNES,KSP,NPC,PC,DM,Mat,Vec,VecScatter,IS,PetscSF,PetscSection,PetscObject,PetscInt,PetscScalar,PetscReal,PetscBool,InsertMode,PetscErrorCode}, 
keywordstyle=\ttb\scriptsize\color{deepblue},
emph={PETSC_COMM_WORLD,PETSC_NULL,SNES_NGMRES_RESTART_PERIODIC},          
emphstyle=\ttb\scriptsize\color{deepred},    
commentstyle=\scriptsize\color{brown},
stringstyle=\ttm\scriptsize\color{deepgreen},
frame=tb,                         
showstringspaces=false            %
}}

\lstnewenvironment{cprog}[1][]
{
\cstyle
\lstset{#1}
}
{}

\newcommand\cinline[1]{{\cstyle\lstinline!#1!}}

\newcommand\bashstyle{\lstset{
language=bash,
basicstyle=\scriptsize\ttfamily,
}}

\lstnewenvironment{bash}[1][]
{
\bashstyle
\lstset{#1}
}
{}

\newcommand\bashinline[1]{{\bashstyle\lstinline!#1!}}

\newcommand\cppstyle{\lstset{
language=C++,
basicstyle=\ttm\scriptsize,
otherkeywords={MPI_Comm,TS,SNES,KSP,PC,DM,Mat,Vec,VecScatter,IS,PetscSF,PetscSection,PetscObject,PetscInt,PetscScalar,PetscReal,PetscBool,InsertMode,PetscErrorCode}, 
keywordstyle=\ttb\scriptsize\color{deepblue},
emph={PETSC_COMM_WORLD,PETSC_NULL,SNES_NGMRES_RESTART_PERIODIC},          
emphstyle=\ttb\scriptsize\color{deepred},    
commentstyle=\scriptsize\color{brown},
stringstyle=\ttm\scriptsize\color{deepgreen},
frame=tb,                         
showstringspaces=false            %
}}

\lstnewenvironment{cpp}[1][]
{
\cppstyle
\lstset{#1}
}
{}

\newcommand\fortranstyle{\lstset{
language=Fortran,
basicstyle=\ttm\scriptsize,
otherkeywords={MPI_Comm,TS,SNES,KSP,PC,DM,Mat,Vec,VecScatter,IS,PetscSF,PetscSection,PetscObject,PetscInt,PetscScalar,PetscReal,PetscBool,InsertMode,PetscErrorCode}, 
keywordstyle=\ttb\scriptsize\color{deepblue},
emph={PETSC_COMM_WORLD,PETSC_NULL,SNES_NGMRES_RESTART_PERIODIC},          
emphstyle=\ttb\scriptsize\color{deepred},    
commentstyle=\scriptsize\color{brown},
stringstyle=\ttm\scriptsize\color{deepgreen},
frame=tb,                         
showstringspaces=false            %
}}

\lstnewenvironment{fortran}[1][]
{p
\fortranstyle
\lstset{#1}
}
{}

\newcommand\makestyle{\lstset{
language=Make,
basicstyle=\ttm\scriptsize,
otherkeywords={MPI_Comm,TS,SNES,KSP,PC,DM,Mat,Vec,VecScatter,IS,PetscSF,PetscSection,PetscObject,PetscInt,PetscScalar,PetscReal,PetscBool,InsertMode,PetscErrorCode}, 
keywordstyle=\ttb\scriptsize\color{deepblue},
emph={PETSC_COMM_WORLD,PETSC_NULL,SNES_NGMRES_RESTART_PERIODIC},          
emphstyle=\ttb\scriptsize\color{deepred},    
commentstyle=\scriptsize\color{brown},
stringstyle=\ttm\scriptsize\color{deepgreen},
frame=tb,                         
showstringspaces=false            %
}}

\lstnewenvironment{make}[1][]
{
\makestyle
\lstset{#1}
}
{}

\title{The Landau Collision Integral in the Particle Basis in the PETSc Library}
\author{Joseph V. Pusztay, Filippo Zonta,  Matthew G. Knepley, Mark F. Adams}
\begin{document}
\maketitle
\begin{abstract}
  The Landau collision integral is often considered the gold standard in the context of kinetic plasma simulation due to its conservative properties, despite challenges involved in its discretization. The primary challenge when implementing an efficient computation of this operator is conserving physical properties of the continuum equation when the system is discretized. Recent work has achieved continuum discretizations using the method of Finite Elements which maintain conservation of mass, momentum, and energy, but which lacks monotonic entropy production. More recently, a particle discretization has been introduced which conserves mass, momentum, and energy, but maintains the benefit of monotonic entropy production necessary for the metriplecticity of the system. We present here an implementation of the particle basis Landau collision integral in the Portable Extensible Toolkit for Scientific Computing in 2 and 3V for the construction of a full geometry solver with a novel approach to computation of the entropy functional gradients. Verification of the operator is achieved with thermal equilibration and isotropization tests. All examples are available, open source, in the PETSc repository for reproduction.
\end{abstract}
\cleardoublepage
\section{Introduction}
The metriplectic representation of a kinetic plasma ~\cite{Morrisondoi:10.1063/1.4982054, MORRISON1986410} poses the system in terms of two brackets; the Poisson bracket where Hamiltonian dynamics may be evolved symplecticly, and the metric bracket which evolves the system according to collisional (dissipitive) effects. This system, by construction, has favorable mathematical properties such as conservation of mass, momentum, and energy of the modified Hamiltonian in the symplectic structure, as well as the monotonic production of entropy in the dissipitive bracket. Discretization of the continuum in the Poisson bracket is relatively straightforward and has been well studied ~\cite{Morrisondoi:10.1063/1.4982054} with conservative time stepping schemes well understood~\cite{Hairer, hairer_lubich_wanner_2003}, where conservative discretizations of the metric bracket require successful discretization of some collision operator. The Landau collision integral \cite{LandauLD1958TKEI} is often considered the gold standard in this regard.

Extensive efforts have been expended on the Landau collision operator, casting it in terms of the metric bracket \cite{doi:10.1063/1.4998610, HirvijokiKrausBurby2018} for continuum and particle representations. Recent work has discretized the operator in its finite element representation whilst maintaining conservation~\cite{contlandau}, which was implemented in PETSc \cite{petsc-user-ref, petsc-web-page, abhyankar2018petsc}, and demonstrates decent performance at exascale ~\cite{AdamsBrennanKnepleyWang2021, adams2023performance}. For particle codes, this demands a projection operation between particle and grid representations, and then a projection of the evolved distribution function weights back to the particles. Suitable, scalable operators exist within PETSc~\cite{conservativeProjection} for this purpose, however there are caveats to this method. Firstly, the necessity of projection to a grid in codes that are purely particle based. The second being the lack of monotonic entropy production in the finite element discretization. A natural remedy to the first is to construct a collision operator which exists purely in the particle basis, which recent theoretical advancements have made possible in a manner that alleviates the second point of concern.

Carillo et. al~\cite{carilloplandau} recently proved the Maxwellian steady state of a regularized particle basis representation of the Landau collision integral, as well as the conservation of mass, momentum, and energy. This discretization however suffered from lack of conservation in energy without a suitable time discretization. A solution to this problem was presented by Hirvijoki in~\cite{eeroparticlelandau}, with a simple conservative time discretization alongside a discrete gradient time stepping scheme which are conservative in mass, momentum, and energy, as well as monotonic in entropy production. Tests of these conservative properties were additionally performed in \cite{zonta} with a GPU implementation of this operator in 2 dimensions. These efforts form the necessary foundation to develop a rigorous community implementation of the particle basis Landau collision integral for the construction of, or implementation into, general Particle-in-Cell (PIC) codes with support for full geometry solvers, with potential for an implementation of the Discrete-Gradient operator.

We present here an implementation of the particle basis Landau collision integral with the time discretization presented in ~\cite{eeroparticlelandau} with 2 particle species of realistic mass ratios in 2 and 3 dimensions. To begin, we will reiterate the metriplectic formulation of the Vlasov-Poisson-Landau system alongside the new particle basis approach to its evaluation. We present a novel approach to evaluation of the entropy functionals via high order quadrature on implicitly defined domains\cite{SAYE2022110720} and compare the quality of such methods against the standard Gauss-Hermite quadratures used in \cite{zonta}. Verification of the methods are demonstrated via basic tests such as thermal equilibration and isotropization of multiple species.
\section{Landau Collisions in the Particle Basis}\label{sec:landau}
Consider the evolution of a \textit{distribution function}, $f$, with respect to time. In a system of charged particles with self consistent and external forces, this reads as
\begin{align}\label{Vlasov_Maxwell_Boltzmann}
    \frac{\partial f}{\partial t} &= \frac{\partial f}{\partial t} + \frac{\partial \Vec{x}}{\partial t} \cdot \nabla_x f + \frac{\partial \Vec{v}}{\partial t} \cdot \nabla_v f \nonumber \\
    &= \frac{\partial f}{\partial t} + \Vec{v} \cdot \nabla_x f + \frac{e}{m} (\Vec{E}+\Vec{v}\cross \Vec{B}) \cdot \nabla_v f \nonumber \\
    &= C
\end{align}
which is the Vlasov-Maxwell-Boltzmann model with velocity $\Vec{v}$, position $\Vec{x}$, electric field $\Vec{E}$, magnetic field $\Vec{B}$, mass $m$, charge $e$, and diffusive collisional term C. This system contains two primary constituents; namely the Poisson bracket corresponding to the electromagnetic terms with $\frac{\partial f}{\partial t} = 0$,  and the diffusive metric bracket incorporating collisional terms as
\begin{align}
    \frac{\partial f}{\partial t} &= \frac{e}{m} \Vec{E}\cdot \nabla f \nonumber \\
    &= C
\end{align}
where C is expressed in the continuum as the Landau collision integral
\begin{align}\label{eq:Landau}
    C = \nu_{\alpha\beta} \frac{m_0}{m_\alpha}\int_\Omega U(\Vec{v}, \bar v)\cdot (\frac{m_0}{m_\alpha}\bar f_\beta \nabla f_\alpha - \frac{m_0}{m_\beta} f_\alpha \bar\nabla \bar f_\beta) 
\end{align}
for species $\alpha$ and $\beta$.

Assume now we have a PIC discretization of the distribution function, $f$, with the distribution function sampled via Dirac delta functions. This generates a phase space distribution of the form 
\begin{align}\label{eq:fullDistributionFunction}
  f = \sum_p w_p \delta(z-z_p)
\end{align}
where p represents the particle indices, $w$ the particle weights, and the configuration space variable $z_p=(x_p, v_p)$. It is clear that this basis doesn't allot evaluation of the collision operator of \ref{eq:Landau} due to \ref{eq:fullDistributionFunction} being neither smooth nor differentiable. One option is to deposit particles to the grid and evaluate the collision integral with finite elements and synethesize the new weights from the grid at the particles. An in depth conservative discretization of such a method is given in ~\cite{contlandau} with accuracy and performance benchmarks in~\cite{AdamsBrennanKnepleyWang2021} and the conservative projection procedure discussed in ~\cite{conservativeProjection}. A more natural approach for codes which do not inherently employ a grid based method would be to formulate \ref{eq:Landau} such that the basis of particles are represented sufficiently smooth. Recent theoretical advances have proven the necessary qualities of such a formualation ~\cite{carilloplandau}.

Recall the evolution in terms of the brackets
\begin{align}
    \frac{dU}{dt} = \{U,F\} + (U,F)
\end{align}
with $F = H - S$ representing the free energy, and $S$ the entropy. The first term, $\{U, F\}$ represents Hamiltonian dynamics in the Poisson bracket, with the right term, $(U,F)$ representing collisional dynamics in the metric bracket. The mass, momentum, and energy are Casimirs of both brackets with the Hamiltonian being a Casimir of the Poisson and metric bracket. The entropy functional is a Casimir of the Poisson bracket, but the metric bracket enforces $\frac{dF}{dt}\geq 0$ corresponding to monotonic entropy production. For an in depth discussion on these Casimir invariants, see ~\cite{doi:10.1063/1.4998610, HirvijokiKrausBurby2018,Morrisondoi:10.1063/1.4982054}.

We can think of the collision integral as advective by introducing the entropy functional
\begin{align}\label{eq:contEntropyFunctional}
    S[f] = -\int f ln f
\end{align}
and using the identity
\begin{align}
  \partial_v f = f\partial_v ln f = - f \partial_v \frac{\delta S}{\delta f}
\end{align}
with 
\begin{align}
    \frac{\partial f}{\partial t} + \frac{\partial  f}{\partial v} \cdot (Ff) = 0
\end{align}
and 
\begin{align}\label{eq:advectiveLandau}
    F[f](v_p) = \gamma \int f(v')Q(\xi) \cdot (\frac{\partial}{\partial v_p}\frac{\delta S}{\delta f} - \frac{\partial}{\partial v_{\bar p}} \frac{\delta S}{\delta f})
\end{align}
Hirvijoki demonstrates for arbitrary functionals, we have the identity
\begin{align}\label{eq:frechetIdentity}
    \frac{\partial}{\partial v_p} \frac{\delta A}{\delta f} = \frac{1}{w_p} \frac{\partial A}{\partial v_p}
\end{align}
for distributions, $f$, of the form in \ref{eq:fullDistributionFunction}, and
\begin{align}
    Q(\xi) = \frac{1}{|\xi|^2} (I - \frac{\xi^T\xi}{|\xi|})
\end{align} 
being the scaled projection matrix with
\begin{align}
    \xi =  v_p - v_{\bar p}
\end{align}
This formulation demonstrates the evolution of the particle velocities in terms of the entropy functional, \ref{eq:contEntropyFunctional}, but the caveat remains the lack of continuity in $f$.

The remaining component for the evaluation of this operator comes in the form of a mollification procedure  ~\cite{carilloplandau} with the mollification process in ~\cite{carilloblobmethod}. Recall again the entropy functional \ref{eq:contEntropyFunctional} over the particle distribution. We can generate a smooth, continuous distribution function in terms of particles by convolution with some smooth radial basis function. For instance, for a Gaussian given by
\begin{align}
    \psi_\epsilon (v-v_p) = \frac{1}{2\epsilon} e^{-(v-v_p)^2/\epsilon}
\end{align}
such that
\begin{align}
    S_\epsilon = \int \psi_\epsilon * f ln \psi_\epsilon * f
\end{align}
In practice, this results in replacing the delta functions in the evaluation of the entropy functional with Gaussians regularized by the parameter $\epsilon$, controlling how "wide" particle interactions are in terms of its entropy. The resultant particle basis entropy functional reads
\begin{align}\label{eq:mollifiedEntropy}
    S_\epsilon = \sum_{v_p} \int \psi_\epsilon(v-v_p) ln \sum_{v_{\bar p}} \psi_\epsilon(v-v_{\bar p})
\end{align}
Now we can express the evolution of the velocity of a particle in terms of the entropy functional gradients around that particles as
\begin{align}\label{eq:timeContLandau}
    \frac{dv_p}{dt} = \sum_{v_{\bar p}}1(p, \bar p) Q(\xi) \Gamma(S_\epsilon, p, \bar p)
\end{align}
where we have maintained Hirvijoki's convention of $1(p, \bar p)$ evaluating to 1 if particles are within the same collision cell, and zero otherwise, and the term $\Gamma(A, p, \bar p) = \frac{1}{w_p}\frac{\partial A}{\partial v_p} - \frac{1}{w_{\bar p}}\frac{\partial A}{\partial v_{\bar p}}$ from the identity in \ref{eq:frechetIdentity}. 

In the first paper, ~\cite{carilloplandau}, the mathematical rigor was focused primarily on proving the Maxwellian steady state and conservative properties of this integral and forward Euler was used for the numerical tests. This time discretization was not conservative in energy, and the solution was put forward by Hirvijoki in ~\cite{eeroparticlelandau} where he evaluates the scaled projection matrix in terms of half time steps to maintain the conservative properties of the time continuous form given in \ref{eq:timeContLandau}. This time discretization reads
\begin{align}\label{eq:particleLandauTimeStep}
    \frac{v_p^{n+1} - v_p^n}{\Delta t} = \sum_{v_{\bar p}} Q(\xi^{(n+\frac{1}{2})}) \Gamma(S_\epsilon^n, p, \bar p)
\end{align}
and grants the conservation of mass, momentum, and energy.
The final component remains the evaluation of the gradient with respect to \ref{eq:mollifiedEntropy} as the $\Gamma(S, p, \bar p)$ term requires it in relation to particle velocity at index p. The derivation is straightforward so long as we are careful of particle indices.
\begin{align} \label{eq:entropyGrad}
    \frac{\partial S_\epsilon}{\partial v_p} &= -\int \nabla_{v_p} \sum_p w_p \psi_\epsilon (v-v_p)ln\sum_k w_k \psi_\epsilon(v-v_k) \nonumber&&\\
    &= -\int \sum_p w_p \psi^{'}(v-v_p) \delta_{pq} ln \sum_k w_k \psi(v-v_k) \nonumber&&\\
    & -\int \sum_q w_q \psi(v-v_q)\frac{1}{\sum_k w_k \psi(v-v_k)}\sum_k w_k \psi(v-v_k)\delta_{kp}\nonumber&&\\
    &= -\int w_p \psi^{'}(v-v_p) ln \sum_k w_k \psi(v-v_k) \nonumber&&\\ 
    & -\int \sum_q w_q \psi(v-v_q) \frac{w_p\psi'(v-v_p)}{\sum_k w_k \psi(v-v_k)}\nonumber&&\\
    &=-\int w_p \psi'(v-v_p) ln \sum_k w_k \psi(v-v_k) - \int w_p \psi'(v-v_p)\nonumber&&\\
    &=-\int w_p \psi'(v-v_p)(1 + ln \sum_k w_k \psi(v-v_k)) 
\end{align}
An important point to note before moving on to numerical implementation and testing is what the equation in \ref{eq:particleLandauTimeStep} is evolving. This formulation results in an evolution of the particle state in phase space (acts directly on velocity), whereas classic interpretations of the integral capture the transfer of momentum through redistribution of the particle weights according to the evolution of the distribution function on the grid.

Another nuance here is the sampling regime these methods converge for, and the importance of the parameter $\epsilon$, as this has stringent implications on the convergence of the particle method to physically meaningful solutions. For a detailed discussion on the convergence of the methods, see \cite{carilloblobmethod}, however it is important enough to briefly discuss again here. 

With the particle spacing
\begin{align}
    h = \frac{2L}{N}
\end{align}
with N being the number of particles along any given slice of an axis and total number of particles given by $N^d$. This gives and expression for $\epsilon$ which has been experimentally demonstrated to converge well as $\epsilon = 2h^{1.98}$. These grid methods typically converge with $\epsilon = o(h)$ \cite{carilloplandau}. This, for the moment, precludes an expression for $\epsilon$ in terms of the discrepency of the set used to sample $f$. In other words, a Klimontovich representation with Monto-Carlo or quasi Monte-Carlo initializations of $f$ with evenly distributed weights are infeasible with this expression. An $\epsilon$ can be found to produce physical results in this regime, but not reliably in experimentation. An expression cast in terms of the discrepency is an active area of research by the authors.
\section{Numerical Implementation and Experiment}\label{sec:pic}
We present in this section the details on the implementation of the collision operator and some numerical experiments. To start, we will discuss the details of time discretization as well as how the quadrature is handled to evaluate the velocity space gradients of \ref{eq:contEntropyFunctional} in \ref{eq:particleLandauTimeStep}. All code for these examples is publicly available in the main branch of PETSc (at the time of writing, version 3.19).
\subsection{Numerical Tools}
It is obvious by the form of \ref{eq:particleLandauTimeStep} in relation to the entropy functional that the particle loop constitutes an $O(N^2)$ computation per time step, necessitating careful handling. This is further compounded by the implicit solve necessary to compute the particle evolution at each time step, or fixed point method used in \cite{zonta}. For this task, PETSc's time stepping library is used to construct a sufficiently accurate and fast solver such that we can evaluate collisions in a reasonable amount of time, which are additionally straightforward to parallelize by construction. 

To that purpose, the main particle loop is handled via Implicit Midpoint to evaluate the $Q(\xi^{n+\frac{1}{2}})$, whereas $\Gamma(S, p, \bar p)$ terms are evaluated at time $n$ and are thus precomputed prior to each time step to reduce overhead in each solver iteration. The Jacobian is approximated using either a standard finite difference, or matrix free approach. The matrix free approach drastically improves runtime, particularly due preallocations of the Jacobian matrix being nonoptimal at the time of writing. Particle management is performed by the PETSc object DMSwarm, which initializes the distribution functions, and provides data structure for particle management as well as algorithms for particle migration between MPI ranks in parallel runs.

The initial particle distribution is the regular sampling scheme. Here, particles are laid out in a grid with the spacing determined by $h = \frac{2L}{N}$, where $[-L,L]^d$ is the domain represented by the grid, and N the number of particles along a line in one of the dimensions. Particle weights are assigned by performing the integral 
\begin{align}
    w_p = \int_{v_p - h/2}^{v_p + h/2} \frac{1}{\sqrt{2\pi\theta}} e^{(-\Vec{v_p}^2/\theta)}
\end{align}
with $\theta = k_BT_s/m_s$ in the isotropic case. Note that the spatial component is omitted as this exists within one spatial cell. Extension to a global maxwellian normalized to all space within spatial cells would merely require regularization of the particle number to the global sum
\begin{align}
    f_p = \sum_p w_p \delta(v-v_p) = 1
\end{align}
extension to which is trivial in this scheme for the collision operator.

\subsubsection{Quadrature}
Given the form of \ref{eq:entropyGrad}, a sufficiently accurate quadrature scheme is necessary to perform the integration. This can naturally be a Gauss-Hermite quadrature given the form of the entropy functional, however we use a different technique. Saye presents an approach to generating high order qudratures on implicit domains defined by polynomials in ~\cite{SAYE2022110720}, which is publicly available in the package Algoim. The benefits of using such a package are the ability to tailor a high order quadrature for the computational domain of varying degree and fidelity.

\begin{figure}[tbhp]
    \label{fig:grid}
    \centering
    \subfloat[Planar Quadrature]{\label{fig:quad2D}\includegraphics[width=.5\textwidth]{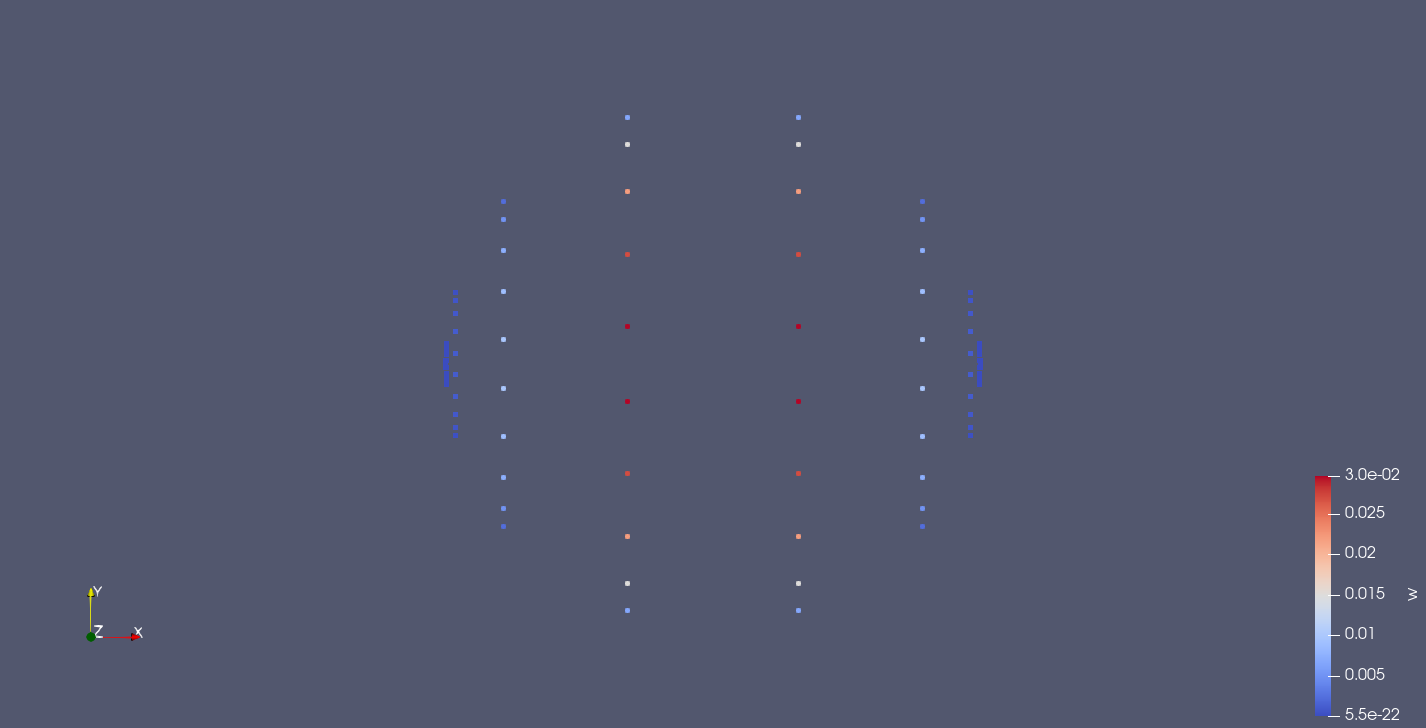}}
    \subfloat[Spherical Quadrature]{\label{fig:quad3D}\includegraphics[width=.5\textwidth]{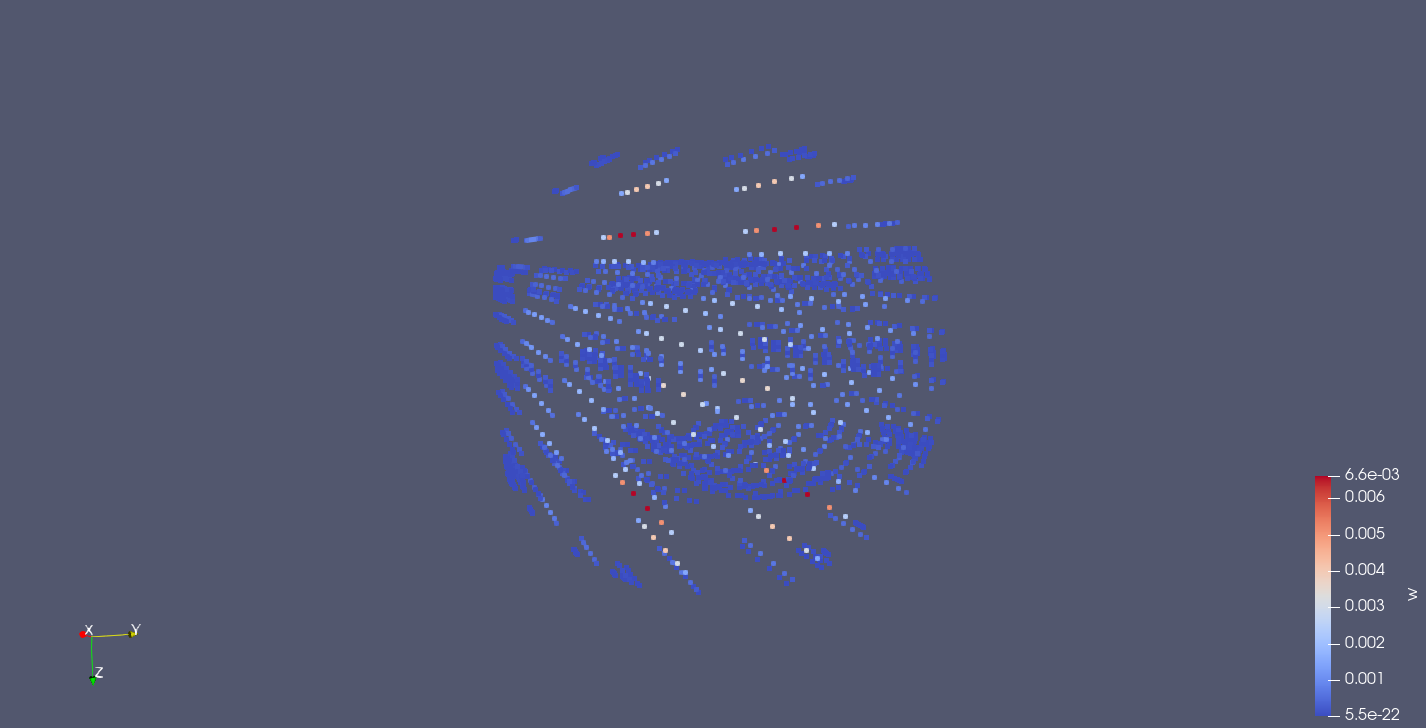}}\\
    \caption{Quadratures in 2 (left) and 3 (right) dimensions generated by algoim. Quadratures are third order with 6 points per level set.}
    \label{fig:quadrature}
\end{figure}

To compute integrals over $S_\epsilon$, we generate a high order quadrature over a disc in 2V, and a ball in 3V. The domain is selected to sufficiently encapsulate the particles within the quadrature. Figure \ref{fig:quadrature} presents a visualization of these quadrature schemes in Paraview. Domains are updated and new quadratures formed as $\epsilon$ evolves. This ensures, without making a change of variables at each step, that the quadrature accurately encapsulates evaluations of the entropy functional gradients.

\subsection{Equilibration}
The most basic test for the collision operator is the appropriate thermal convergence of two thermal populations. The first population is given a temperature of 0.35KeV, whereas the second population is given a temperature of 0.2KeV. The distribution then relaxes to an equilibrium state. In this case, an initial electron-positron distribution is selected to demonstrate conservative properties of the operator in a trivial case of $m_1/m_2 = 1$. Larger mass ratios are investigated in later tests such as isotropization. 

\begin{figure}[tbhp]
    \centering
    \subfloat[Equilibration with Gauss-Hermite quadrature. ]{\label{fig:GH_equilibration}\includegraphics[width=.5\textwidth]{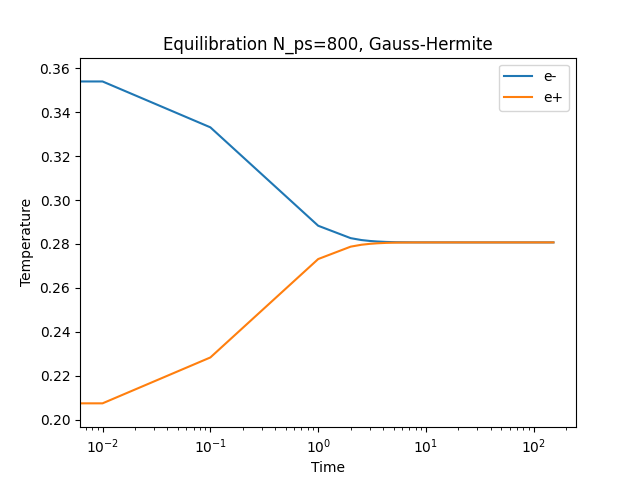}}
    \subfloat[Equilibration with Algoim]{\label{fig:temps}\includegraphics[width=.5\textwidth]{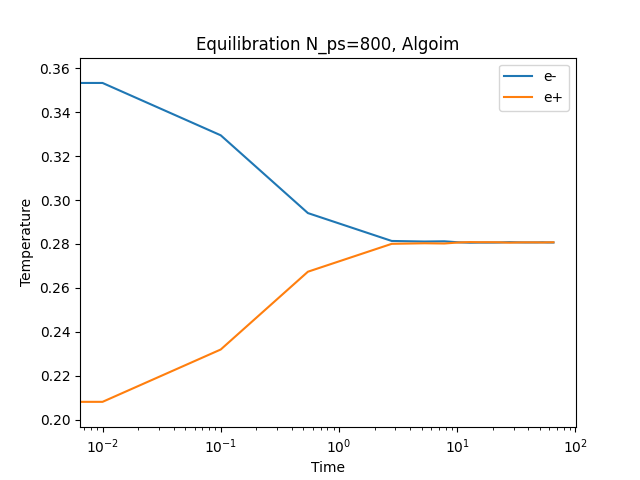}}\\
    \subfloat[Gauss-Hermite relative error in kinetic energy ]{\label{fig:GH_equilibration}\includegraphics[width=.5\textwidth]{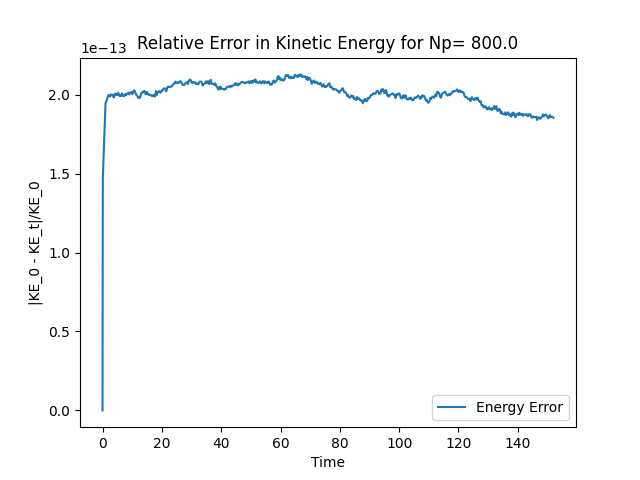}}
    \subfloat[Algoim relative error in kinetic energy]{\label{fig:temps}\includegraphics[width=.5\textwidth]{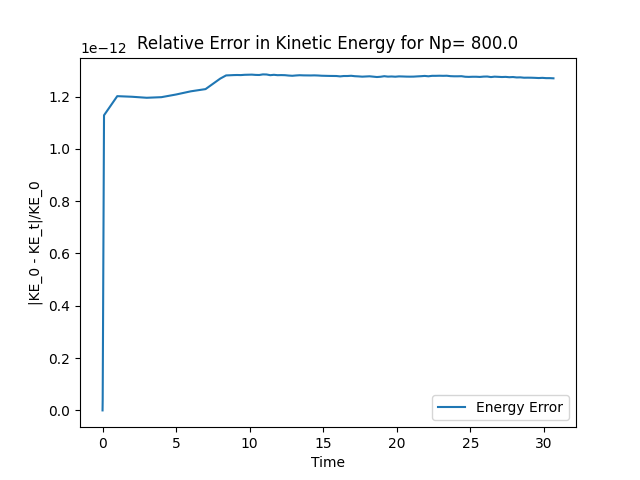}}\\
    \caption{Convergence to equilibrium for electron-positron collisions with both Gauss-Hermite and Algoim evaluations of the entropy functional gradient quadrature. An initial jump in energy error occurs on the order of $10^-12$ and remains consistent with numerical noise.}
    \label{fig:tems_2d}
\end{figure}
The appropriate behavior is observed with the two populations effectively reaching an equilibrium state, with error on the order of $10^{-12}$ in kinetic energy of the system. Effective error reduction can be achieved by utilizing a smaller time step, as well as using the dense finite difference approximation to the Jacobian. The matrix free approach is leveraged here with relatively large time steps, where the solver is easily capable of converging. This is not the case with species presenting relatively large mass ratios on the order of realistic electron-ion collisions, where the step size must be significantly reduced. This test case is also light on the quadrature used, with a 3rd order quadrature generated over the disc using 10 points per level set.

\subsection{Two Species Isotropization}
In isotropization tests, each velocity space dimension is given a different value for its temperature for each species in 2V. In the 3V case, we have $T_x = T_y$ to generate a uniform parallel and transverse temperature. The relaxation rate is given in the Naval Research Laboratory Plasma Formulary (NRLPF) as ~\cite{nrlformulary} 
\begin{align}
    \frac{dT_\perp}{dt} = \frac{1}{2}\frac{dT_{||}}{dt} = -v_T^\alpha (T_\perp - T_{||})
\end{align}
for $A= T_\perp / T_{||} - 1 > 0$ we have 
\begin{align}
    v_T^\alpha = \frac{2\sqrt{\pi}e_\alpha^2 e_\beta^2 n_\alpha \lambda_{\alpha \beta}}{m_\alpha^{1/2}(kT_{||})^{1/2}}A^{-2}[-3+(A+3)\frac{tan^{-1}(A^{1/2})}{A^{1/2}}]
\end{align}
When $A < 0$, the arctan is replaced with hyperbolic arctan and its arguments negated.
Additional parameters include the mass, in which case we use the electron mass for species 0, and two times the proton mass for species 1 (an approximate Deuteron). Isotropization results are shown to agree well with empirical results from the NRL.

\begin{figure}[tbhp]
    \centering
    \subfloat[Isotropization with Gauss-Hermite quadrature]{\label{fig:gh_2v_iso}\includegraphics[width=.5\textwidth]{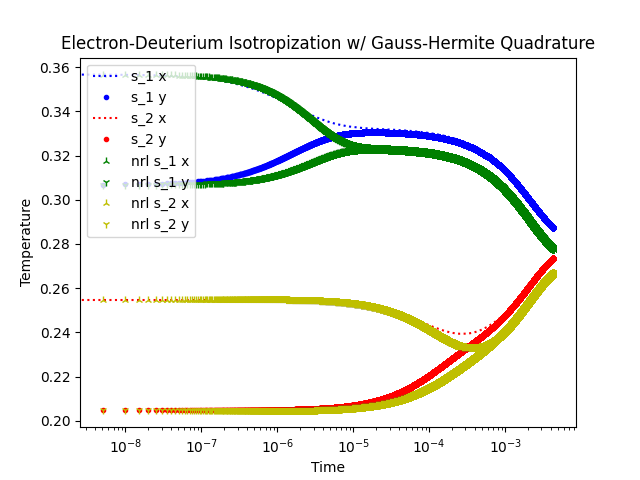}}
    \subfloat[Isotropization with Algoim]{\label{fig:temps}\includegraphics[width=.5\textwidth]{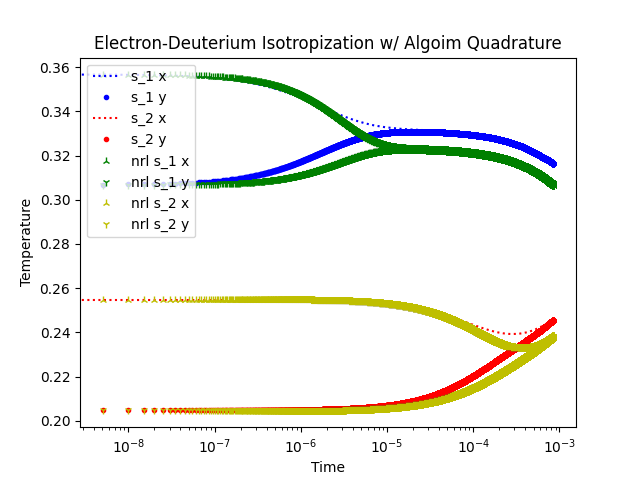}}\\
    \caption{Two species isotropization with quadrature over the disc (right) and gauss hermit quadrature (left) in 2V. Variation in rate from the NRL is attributed to the NRL assuming each successive step is Maxwellian where there is no guarantee in the intermediate steps towards convergence.}
    \label{fig:tems_2d}
\end{figure}
Decent agreement is seen with a minor deviation in the rate. This is attributed to the assumptions in the empirical equations that the intermediate states of the distribution are Maxwellian, where the guarantee of closness to the Maxwellian of intermediate states is not enforced. Adams et al. \cite{adams2023performance} and Hager \cite{Hager2016} present qualitatively similar results.
\section{Concluding Remarks and Future Developments}\label{sec:conclusion}
We have reiterated the metriplectic formulation, and the recent theoretical advancements that have made evaluation of the Landau collision integral possible. Demonstration of the appropriate behavior of the collision integral was performed on common tests for collision operators and compared to data from the NRLPF, providing cross verification of the collision integral done by Zonta et al. in a completely separate implementation, for a different machine architecture, with different methods for computing its integrals. All methods demonstrate considerable agreement. In addition to cross verification, we presented and provided a community software implementation of this collision integral for use in the construction of, or supplementation of, large scale plasma codes. Naturally, some additional studies and optimizations are a subject of ongoing development. 

The stringent structure of the sampling method to obtain stability of this operator is of first consideration. In the case of simulations which use free streaming particles, there is no guarantee of the structure at each collision step being sufficiently regular, as such, two solutions are posed to be investigated, firstly the method of coarse graining. In this case, particles are projected from the input basis to a sufficiently regular particle basis. Weights would then be determined according to the projected moments of the initial basis using the particle mass matrix presented in \cite{conservativeProjection}. The second method, which is more mathematically rigorous, is the cast the epsilon parameter in terms of the discrepency of sampling function. This is a more natural approach but would incur the additional overhead of measuring the discrepency of the sample prior to the collision step. As stated, both of these avenues are left to future work.

The final optimization to be examined is the extension of the PETSc implementation to GPUs, similar to the work performed in \cite{zonta} by Zonta et al, but in the PETSc Vec GPU back end. This work is underway and similarly to the operator provided in the continuum will leverage a Kokkos \cite{CarterEdwards20143202} backend for portability requirements. This additional component sets the stage for real world physics studies and comparisons between collision operators on exascale problems with a particle basis full geometry solver in conjunction with additional verification test of the Discrete Gradients implementation underway in PETSc as well.
\section{Acknowledgements}
This work was supported in part by the U.S. Department of Energy, Office of Science, Office of Fusion Energy Sciences and Office of Advanced Scientific Computing Research, Scientific Discovery through Advanced Computing (SciDAC) program through the FASTMath Institute under Contract No. DE-AC02-05CH11231 at Lawrence Berkeley National Laboratory. Additionally,  this work was funded in part by the Exascale Computing Project (17-SC-20-SC), a collaborative effort of the U.S. Department of Energy Office of Science and the National Nuclear Security Administration. We gratefully acknowledge assistance on the Geosolver cluster at the UB Center for Computational Research by CCR staff, and the author of Algoim, Robert Saye, for insight into effective usage of the package and its algorithms.
\printbibliography
\end{document}